\documentstyle[epsf,12pt]{article}
\hoffset = - 1 truecm
\def\beq{\begin{equation}}   \def\eeq{\end{equation}}
\def\bea{\begin{eqnarray}}  \def\eea{\end{eqnarray}} \def\nn{\nonumber}
\def\noi{\noindent} \def\beeq{\begin{eqnarray}}
\def\eeeq{\end{eqnarray}}
\def\lsim{\raise0.3ex\hbox{$<$\kern-0.75em\raise-1.1ex\hbox{$\sim$}}}
\def\gsim{\raise0.3ex\hbox{$>$\kern-0.75em\raise-1.1ex\hbox{$\sim$}}}

\newcommand\mysection{\setcounter{equation}{0}\section}
\renewcommand{\theequation}{\thesection.\arabic{equation}}
\newcounter{hran} \renewcommand{\thehran}{\thesection.\arabic{hran}}

\def\bmini{\setcounter{hran}{\value{equation}}
\refstepcounter{hran}\setcounter{equation}{0}
\renewcommand{\theequation}{\thehran\alph{equation}}\begin{eqnarray}}

\def\bminiG#1{\setcounter{hran}{\value{equation}}
\refstepcounter{hran}\setcounter{equation}{-1}
\renewcommand{\theequation}{\thehran\alph{equation}}
\refstepcounter{equation}\label{#1}\begin{eqnarray}}

\def\emini{\end{eqnarray}\relax\setcounter{equation}{\value{hran}}\renewcommand{\theequation}{\thesection.\arabic{equation}}}

\def\ben{\begin{enumerate}}  \def\een{\end{enumerate}}

\def\cite#1{[\ref{#1}]} \def\citd#1#2{[\ref{#1}, \ref{#2}]}
\def\citt#1#2#3{[\ref{#1}, \ref{#2}, \ref{#3}]}
\def\citm#1#2{[\ref{#1}--\ref{#2}]}

\begin{document} 
\begin{center} 
\vbox to 1 truecm {} {\Large \bf The Cosmological Constant}
\footnote{Lecture given at the XIV Workshop "Beyond the Standard
Model", Bad Honnef, 11-14  March 2002}
\vskip 1 truecm {\bf U. Ellwanger} \vskip 3 truemm

{\it Laboratoire de Physique Th\'eorique\footnote{Unit\'e Mixte de
Recherche CNRS - UMR N$^{\circ}$ 8627},\\ Universit\'e Paris XI,
B\^atiment 210, 91405 Orsay Cedex, France\\ E-mail:
Ulrich.Ellwanger@th.u-psud.fr} \end{center} 
\vskip 1.5 truecm

\centerline{\bf  Abstract}
\vskip 5 truemm
Various contributions to the cosmological constant are discussed and
confronted with its recent measurement. We briefly review different
scenarious -- and their difficulties -- for a solution of the
cosmological constant problem.

\vskip 2 truecm

\centerline{\bf Contents}
\vskip 5 truemm

\hskip 2 truecm 1) Friedmann-Robertson-Walker Cosmology \par

\hskip 2 truecm 2) Measurement of the Cosmological Constant \par

\hskip 2 truecm 3) The Cosmological Constant in Classical and Quantum 
Field Theory \par

\hskip 2 truecm 4) Towards Solutions of the Problem: \par

\hskip 2 truecm \qquad a) Supersymmetry \par
\hskip 2 truecm \qquad b) Dilaton \par
\hskip 2 truecm \qquad c) 3-Form Field \par
\hskip 2 truecm \qquad d) Quintessence \par
\hskip 2 truecm \qquad e) Brane Universes \par

\hskip 2 truecm 5) Conclusions \par 
\vskip 1.5 truecm

\noindent LPT Orsay 02-18 \par
\noindent March 2002\par

\newpage \pagestyle{plain} \baselineskip=18 pt

\mysection{Friedmann-Robertson-Walker Cosmology}
\hspace*{\parindent}The most common physical interpretation of General
Relativity \citm{1r}{3r}\footnote{The following section can not replace
an introduction to General Relativity; to this end see the
corresponding literature.} is that space-time has to be considered as a
generally non-trivial four-dimensional Riemannian manifold: Let
$x^{\mu} \equiv \{x^i, t\}$ be $3 + 1$ coordinates on this manifold,
then there exists a metric $g_{\mu\nu}(x)$ which defines a line element

\beq \label{1.1e} ds^2 = g_{\mu\nu}(x) \ dx^{\mu} \ dx^{\nu} \ . \eeq

\noi Flat Minkowski space corresponds to

\beq \label{1.2e} g_{\mu\nu}(x) = \eta_{\mu\nu} = {\rm diag} (1, -1,
-1, -1) \ . \eeq

\noi An essential feature of General Relativity is that $g_{\mu\nu}(x)$
is considered as a dynamical field which is determined by its equations
of motion, the so-called Einstein equations. They involve the Ricci
tensor $R_{\mu\nu}$ and the Ricci scalar $R$ constructed from the
metric (\ref{1.2e}):

\beq \label{1.3e} R_{\mu\nu} - {1 \over 2} \ g_{\mu\nu} \ R = - \kappa
\ T_{\mu\nu} \eeq

\noi with

\bea \label{1.4e} &&\kappa = 8 \pi {\cal G}/c^2 \cong 1,865 \cdot
10^{-29} \ {\rm m/g} \ , \nn \\ &&{\cal G} = \hbox{Newton's constant
$\cong 6,668 \cdot 10^{-14}$ m$^3$/(g sec$^2$) \ .} \eea

\noi $T_{\mu\nu}$ in (\ref{1.3e}) is the energy-momentum tensor of
matter which acts as source for the gravitational field, the metric
$g_{\mu\nu}(x)$. Its role is analogous to the one of electro-magnetic 
currents
$J_{\mu}$, which act as sources for the electro-magnetic field
$A_{\mu}$ in Maxwell's equations. Note, however, that the left-hand
side of (\ref{1.3e}) is highly nonlinear in $g_{\mu\nu}$ (in some
analogy to non-abelian Yang-Mills field equations), and that also
$T_{\mu\nu}$ depends in general on $g_{\mu\nu}$. This becomes clear if
we derive (\ref{1.3e}) from an action principle. To this end we have to
assume that the equations of motion for matter can be derived from
a Lagrangian ${\cal L}_M$, which can correspond to point particles or
classical fields. For consistency (the Bianchi identities for the
Riemann tensor which imply, {\it via} (1.3), that $T_{\mu\nu}$ is
covariantly conserved) it is necessary that the space-time integrated
matter Lagrangian is invariant under general coordinate
transformations. A corresponding Lagrangian for a real scalar field
$\varphi (x)$, e.g., is given by

\beq \label{1.5e} \int d^4x \ {\cal L}_M (\varphi , g_{\mu\nu}) = \int
d^4 x \sqrt{-g} \left ( {1 \over 2} \partial_\mu \varphi \ g^{\mu\nu} \
\partial_\nu \varphi - V(\varphi ) \right ) \eeq

\noi where

\beq \label{1.6e} g = \det \left ( g_{\mu\nu} \right ) \qquad
\left ( = -1\
\hbox{for} \ g_{\mu\nu} = \eta_{\mu\nu} \right ) \ . \eeq

In general the matter energy momentum tensor $T_{\mu\nu}$ is given by

\beq \label{1.7e} T_{\mu\nu}(x) = {2 \over \sqrt{-g(x)}} \ {\delta \over
\delta g^{\mu\nu}(x)} \int d^4 x' \ {\cal L}_M (x') \ . \eeq

\noi (The factor $\sqrt{-g}$ on the right-hand side of eq. (\ref{1.5e})
renders the volume integration invariant and is sometimes written
explicitly under the $d^4x'$ integral in eq. (\ref{1.7e})). The general
dependence of $T_{\mu\nu}$ on $g_{\mu\nu}$ is now apparent.\par

The left-hand side of eq. (\ref{1.3e}) can also be derived from an
action, the integrated Einstein Lagrangian

\beq \label{1.8e} \int d^4x \ {\cal L}_E = {1 \over 2 \kappa} \int d^4
x \sqrt{-g} \ R \ . \eeq

The variation of eq. (\ref{1.8e}) with respect to $g^{\mu\nu}$ gives,
up to an integral of a total derivative (which can be important
in the context of brane universes, but which is dropped here),

\beq \label{1.9e} {\delta \over \delta g^{\mu\nu} (x)} \int d^4 x' \
{\cal L}_E(x') = {\sqrt{-g} \over 2 \kappa} \left ( R_{\mu\nu} - {1
\over 2} g_{\mu\nu} R \right ) \ , \eeq

\noi hence eq. (\ref{1.3e}) follows from $2\kappa/\sqrt{-g}$ times the
vanishing variation of 

\beq \label{1.10e} \int d^4x \ {\cal L}_{Tot} = \int d^4x \left ( {\cal
L}_E + {\cal L}_M \right ) \ . \eeq

The following configurations of the metric $g_{\mu\nu}$ are of
particular physical importance: \par

a) The Schwarzschild solution around a point-like source (and the Kerr
solution around rotating point-like sources as pulsars): It
determines the dynamics of astronomical objects (stars, planets,
galaxies etc.) and reproduces, in the non-relativistic and weak field
limit, the Newton potential $M{\cal G}/r$. This solution of Einstein
equations is tested down to scales of ${\cal O}$(1 mm).\par

b) Gravitational waves, which are solutions of the Einstein equations
in the vacuum ($T_{\mu\nu} = 0$) and await their discovery. \par

c) Cosmological configurations, which determine the global geometry of
the universe and its temporal evolution. They are the esssential
subject of this chapter. \par

The search for such configurations of the metric $g_{\mu\nu}(x)$ is
greatly simplified by the assumptions of isotropy of the universe (it
looks the same in all directions) and homogeneity of the universe (no
point is singled out, and physically relevant quantities as the Riemann
scalar $R$ and the left-hand side of eq. (\ref{1.3e}), the Einstein
tensor, do not depend on $x^i$). Experimental evidence for these
assumptions is not overwhelming: Seemingly matter is lumped in our
universe, from stars to galaxies to clusters of galaxies up to large
scale structures of the size of the observed part of the universe.\par

Nevertheless the assumption is that matter can be described by an
approximately $x^i$-independent (but $t$-dependent) energy-momentum
tensor $T_{\mu\nu}(t)$. From isotropy it follows that the only
non-vanishing components of $T_{\mu\nu}(t)$ are

\beq \label{1.11e} T_{00}(t) \ , \ T_{11}(t) = T_{22}(t) = T_{33}(t)
\equiv T_s(t) \ . \eeq

A convenient way to write a cosmological configuration of the metric
$g_{\mu\nu}$ is in terms of the Robertson-Walker line element in
spherical coordinates

\beq \label{1.12e} ds^2 = dt^2 - a^2(t) \left ( {dr^2 \over 1 - kr^2} +
r^2 \left ( d \theta^2 + \sin^2 \theta d\phi^2 \right ) \right ) \ .
\eeq

\noi Here the constant $k$ can be chosen, after an appropriate
rescaling of $r$ and $a(t)$, as $k = 0, \pm 1$. It determines the
global geometry of three-dimensional space, which is flat for $k =
0$, a three-dimensional hypersphere (and hence closed) for $k = 1$, and
hyperbolic (open) for $k = - 1$.  $a(t)$ in (\ref{1.12e}) is a scale
factor of the three-dimensional space, and its time dependence is
determined by the 00 or $ii$ components of Einstein equations
(\ref{1.3e}):

\bminiG{1.13e}
\label{1.13ae}
3 \ {\dot{a}^2 + k \over a^2} = \kappa \ T_{00}(t) \ ,
\eeeq
\beeq
  \label{1.13be}
  - \ {2a \ddot{a} + \dot{a}^2 + k \over a^2} = \kappa \ T_s(t)
\emini

\noi where $\dot{a} = da/dt$. Evidently eqs. (\ref{1.13e}) imply some
relation between $T_{00}$ and $T_s$, which can be written as

\beq \label{1.14e} \dot{T}_{00} + 3 \ {\dot{a} \over a} \left ( T_{00}
+ T_s \right ) = 0 \eeq

\noi and corresponds to the covariant conservation of the
energy-momentum tensor. More properties of $T_{00}$, $T_s$ have to be
derived from ans\"atze for the properties of matter. Traditionally
matter is modelled by a perfect fluid with density $\rho (t)$ and
pressure $p(t)$. Even if stars (or galaxies) would be homogeneously
distributed in our universe this model has its subtle problems:
Evidently $\rho (t)$ and $p(t)$ play the role of effective spatial
averages over ``point-like'' stars or galaxies, and $a(t)$ corresponds
to a spatial average of spatial diagonal components of $g_{\mu\nu}(x)$
(the superposition of a large number of Schwartzschild solutions in the
weak field limit).
However, instead of writing eqs. (\ref{1.13e}) for these spatial
averages, one should take the spatial average of Einstein's equations
(\ref{1.3e}). Due to the non-linear nature of these equations this is
{\it not} the same. In addition, the spatial average of a metric
on which spatial volumina depend is very difficult to define properly.
For more discussions and literature on this subject see \cite{4r}.
\par

Within the perfect fluid model of matter the components of the
energy-momentum tensor take the form

\beq \label{1.15e} T_{00} = \rho (t) + \Lambda \quad , \qquad T_s =
p(t) - \Lambda \eeq

\noi where $\Lambda$ is the so-called cosmological constant. Its
physical origin will be discussed in detail below. \par

The equation of state of a perfect fluid determines a relation $p =
p(\rho )$. Usually one assumes

\beq \label{1.16e} p = w \rho \eeq

\noi where the constant $w$ depends on the microscopic properties of
the fluid. Now eq. (\ref{1.14e}) (energy-momentum conservation) assumes
the simple form

\beq \label{1.17e} \dot{\rho} + 3 \ {\dot{a} \over a} (1 + \omega )
\rho = 0 \ . \eeq

It is instructive to derive expressions for the parameters $w$
and $\Lambda$ in the case where matter is modelled by plane waves of a
scalar field. This model is possibly more realistic in an early hot and
dense phase of the universe, but it also helps to interpret the
physical significance of the above quantities at later epochs. \par

Let us consider the scalar field Lagrangian (\ref{1.5e}), with
$V(\varphi )$ developed up to quadratic order in $\varphi$ around its
minimum $V_0$:

\beq \label{1.18e} \int d^4x \ {\cal L}_M^{(2)}(\varphi , g_{\mu\nu}) =
\int d^4x \sqrt{-g} \left ( {1 \over 2} \partial_\mu \varphi \
g^{\mu\nu} \partial_\nu \varphi - V_0 - {1 \over 2} \ m^2 \ \varphi^2
\right ) \ .\eeq

If one constructs the components of the energy momentum tensor
according to eq. (\ref{1.7e}) the following relation is useful:

\beq \label{1.19} {\delta \over \delta g^{\mu\nu}} \ \sqrt{-g} = - {1
\over 2} \ \sqrt{-g} \ g_{\mu\nu} \ . \eeq

Then one obtains

\bminiG{1.20e} \label{1.20ae} T_{00} = \partial_0 \varphi \partial_0
\varphi - g_{00} \left ( {1 \over 2} \partial_{\mu} \varphi g^{\mu\nu}
\partial_{\nu} \varphi - {m^2 \over 2} \varphi^2 - V_0 \right ) \eeeq
\beeq \label{1.20be} T_{ij} = \partial_i \varphi \partial_j \varphi -
g_{ij} \left ( {1 \over 2} \partial_{\mu} \varphi g^{\mu\nu}
\partial_{\nu} \varphi - {m^2 \over 2} \varphi^2 - V_0 \right ) \ .
\emini

Subsequently we will replace, for simplicity, $g_{\mu\nu}(x)$ by
$\eta_{\mu\nu}$ in $T_{\mu\nu}$; this avoids the need to discuss plane
waves in curved space time. Hence we take $T_{\mu\nu}$ to be static;
this still allows us to obtain expressions for $w$ and $\Lambda$ from
eqs. (1.15) and (1.16). Then eqs. (\ref{1.20e}) (cf. our conventions
(\ref{1.2e})) simplify to

\bminiG{1.21e} \label{1.21ae} T_{00} = {1 \over 2} \left ( \left (
\partial_0 \varphi \right )^2 + \left ( \partial_i \varphi \right )^2 +
m^2 \varphi^2 \right ) + V_0 \eeeq \beeq \label{1.21be} T_{ij} =
\partial_i \varphi \partial_j \varphi + {\delta_{ij} \over 2} \left (
\left ( \partial_0 \varphi \right )^2 + \left ( \partial_i \varphi
\right )^2 + m^2 \varphi^2 \right ) - \delta_{ij} V_0 \emini

\noi Next we replace the fields $\varphi (x^i,t)$ by plane waves with
constant amplitude $\varphi_0$:

\beq \label{1.22e} \varphi (x^i, t) = \varphi_0 \sin (\omega t -
\vec{k} \vec{x}) \eeq

\noi where the equations of motion enforce the dispersion relation

\beq \label{1.23e} \omega^2 = \vec{k}^2 + m^2 \ . \eeq

\noi Equations (\ref{1.21e}) become

\bminiG{1.24e} \label{1.24ae} T_{00} = {1 \over 2} \varphi_0^2 \left (
\left ( \omega^2 + \vec{k}^2 \right ) \cos ^2 (\omega t -
\vec{k}\vec{x}) + m^2 \sin ^2 (\omega t - \vec{k} \vec{x}) \right ) +
V_0 \eeeq \beeq \label{1.24be} T_{ij} = \varphi_0^2 \left ( \left ( k_i
k_j - {\delta_{ij} \over 2} \vec{k}^2 + {\delta_{ij} \over 2} \omega^2
\right ) \cos ^2 (\omega t - \vec{k}\vec{x}) - {\delta_{ij} \over 2}
m^2 \sin^2 (\omega t - \vec{k} \vec{x} \right )  - \delta_{ij} V_0 \ .
\emini

\noi Now we perform two averages: First, due to isotropy, we average
over the angular dependence of the wave vectors $\vec{k}$, using

\beq \label{1.25e} <k_ik_j>_{\varphi, \theta} = {\delta_{ij} \over 3} \
\vec{k}^2 \ . \eeq

\noi Second, due to homogeneity, we average over space using

\beq \label{1.26e} <\cos^2 (\omega t - \vec{k}\vec{x} )>_x \ = \ < \sin^2
(\omega t - \vec{k}\vec{x}>_x \ = {1 \over 2} \ . \eeq

\noi The final expressions for the components of the energy-momentum
tensor are (with the notation $T_s$ for its spacial components, cf.
(\ref{1.11e}), and actually we should have written $<T_{00}>$ and
$<T_s>$)

\bminiG{1.27e} \label{1.27ae} T_{00} = {1 \over 4} \ \varphi_0^2 \left (
\omega^2 + \vec{k}^2 + m^2 \right ) + V_0 \eeeq \beeq \label{1.27be}
T_s = {1 \over 4} \ \varphi_0^2 \left ( \omega^2 - {1 \over 3} \ \vec{k}^2
- m^2 \right ) - V_0 \ . \emini

\noi Comparing with eqs. (\ref{1.15e}) and (\ref{1.16e}) we immediately
find for the cosmological constant

\beq \label{1.28e} \Lambda = V_0 \eeq

\noi and for $w$, the ratio of the pressure to density,

\beq \label{1.29e} w = {\omega^2 - {1 \over 3} \ \vec{k}^2 - m^2 \over
\omega^2 + \vec{k}^2 + m^2} = {\vec{k}^2 \over 3(\vec{k}^2 + m^2)} \eeq

\noi where we have used the dispersion relation (\ref{1.23e}) in the
second step in (\ref{1.29e}). From (\ref{1.29e}) one finds for
radiation, $m^2 = 0$,

\beq \label{1.30e} w_{rad} = {1 \over 3} \ , \eeq

\noi and for non-relativistic matter, $\vec{k}^2 \ll m^2$,

\beq \label{1.31e} w_{nr} \cong 0 \ . \eeq

\noi Sometimes $w$ is not defined by the ratio of pressure and density
as in (\ref{1.16e}), but directly by the ratio of $T_s/T_{00}$. Then,
if the cosmological constant $\Lambda$ dominates over $\rho$ over $p$,
one obtains an effective $w_{\Lambda}$ with

\beq \label{1.32e} w_{\Lambda} = {T_s \over T_{00}} = - 1 \ . \eeq

After this intermezzo on the physical meaning of $w$ in eq.
(\ref{1.16e}) we plug the first of eqs. (\ref{1.15e}) into eq.
(\ref{1.13ae}), which becomes

\beq \label{1.33e} 3 \ {\dot{a}^2 \over a^2} = - {3k \over a^2} +
\kappa \rho (t) + \kappa \Lambda \eeq

\noi and we recall eq. (1.17),

\beq \label{1.34e} \dot{\rho}(t) + 3 \ {\dot{a} \over a} (1 + w) \rho
(t) = 0 \ . \eeq

If the term $\kappa \rho (t)$ dominates the right-hand side of eq.
(\ref{1.33e}) (and $w \not= - 1$), the solution of eqs. (\ref{1.33e})
and (\ref{1.34e}) is

\bminiG{1.35e} \label{1.35ae} a(t) = a_0 \ t^{{2 \over 3}(1+w)} \eeeq
\beeq \label{1.35be} \kappa \rho (t) = {4 \over 3(1 + w)^2} \ t^{-2} \
. \emini

\noi Hence the universe expands, and $\rho (t)$, the matter density,
decreases. \par

If the term $\kappa \Lambda$ dominates the right-hand side of eq.
(\ref{1.33e}) (or, alternatively, $w = - 1$ and $\rho (t)$ = const. $=
\Lambda$) the universe expands exponentially, which corresponds to an
inflationary epoch.\par

The ratio $\dot{a}/a$ {\it today} is called the Hubble
constant $H_0$:

\beq \label{1.36e} \left . H_0 = {\dot{a} \over a} \right |_{today} \ .
\eeq

\noi Apparently our universe expands today, i.e. distant galaxies (with
distances measured in $Mpc \cong 3,1\cdot 10^{22}$~m) move away with
apparent speeds (of the order of km/sec) which are roughly proportional
to $H_0$ times their distance.
Hence one measures $H_0$ in these units,

\beq \label{1.37e} H_0 = h_0 \cdot 100 {{\rm km} \over {\rm sec}}
(Mpc)^{-1} \ , \eeq

\noi with

\beq \label{1.38e} h_0 \sim 0.65 \ . \eeq

\noi The motion of distant galaxies reveals itself in a redshift $z$,
i.e.  the measured wavelengths $\lambda_m$ and frequencies $\nu_m$
differ from the emitted wavelengths $\lambda$ and $\nu$:

\beq \label{1.39e} {\lambda_m \over \lambda} = {\nu \over \nu_m} = 1 +
z \ . \eeq

For a time dependent scale factor $a(t)$ one obtains

\beq \label{1.40e} 1 + z = {a(0) \over a(-t)} \eeq

\noi where $-t$ is the time when the signal was emitted. For the 
light-like distance $D$ (or luminosity distance) of these galaxies 
one finds
with the metric (\ref{1.12e})

\beq \label{1.41e} D = a(0)\int_0^r {dr \over \sqrt{1 - kr^2}} = 
a(0)\int_{-t}^{0} {dt' \over a(t')} \ . \eeq

\noi Here we have used (for light-like distances)

\beq \label{1.42e} 0 = ds^2 = dt^2 - a^2(t) \ {dr^2 \over 1 -
kr^2} \ . \eeq

\noi These results will be used in the next chapter. Let us reconsider
eq. (\ref{1.33e}) today, using (\ref{1.36e})

\beq \label{1.43e} 3H_0^2 = - {3 k \over a(0)^2} + \kappa \rho (0) +
\kappa \Lambda \ . \eeq

\noi It is convenient to define

\beq \label{1.44e} \Omega_M = {\kappa \rho (0) \over 3H_0^2} \quad ,
\quad \Omega_{\Lambda} = {\kappa \Lambda \over 3 H_0^2} \eeq

\noi which turns eq. (\ref{1.43e}) into

\beq \label{1.45e} \Omega_M + \Omega_{\Lambda} - {k \over H_0^2 a^2(0)}
= 1 \ . \eeq

\noi Let us recall that inflation predicts $k = 0$ (a spacially flat
universe), hence in this case

\beq \label{1.46e} \Omega_M + \Omega_{\Lambda}  \mathrel{\mathop =^{
!}} 1 \ . \eeq

\noi This value for $k$ is also compatible with recent measurements of
fluctuations of the Cosmic Microwave Background \cite{cmbr}.
 Note that $\rho (0)$ in (\ref{1.44e}), and hence $\Omega_M$,
includes both baryonic (visible or invisible) and dark matter. The
corresponding split will not concern us here; subsequently we will
assume, however, that the equation of state for matter is such that $w
= 0$ corresponding to nonrelativistic (baryonic and/or dark) matter.

\mysection{Measurement of the Cosmological Constant}
\hspace*{\parindent} It turns out that we could measure $\Omega_M$ and
$\Omega_{\Lambda}$ if we could determine the precise dependence of the
distance of astronomical objects on the redshift $z$, $D(z)$: If we put
$w = 0$ in (\ref{1.34e}) (correspond to nonrelativistic matter) one can
solve eqs. (\ref{1.33e}) and (\ref{1.34e}) for $a(t)$, $\rho (t)$, with
$\dot{a}(0)/a(0) = H_0$ and $\rho (0)$ (or $\Omega_M$) and 
$\Omega_{\Lambda}$ as boundary conditions. The solution for
$a(t)$ can be plugged into eq. (\ref{1.41e}) for the luminosity
distance $D$, and the integral $dt'$ can be written as an integral over
$dz'$ using (\ref{1.40e}). The resulting expression for $D(z)$ reads
(re-installing the speed of light $c$)

\beq \label{2.1e} D(z) = {c(1 + z) \over H_0 \cdot \sqrt{|\lambda
|}} \widehat{\sin} \left ( \sqrt{|\lambda |} \int_0^z {dz' \over
\sqrt{(1 + z')^2 (1 + \Omega_M z') - (2 + z') z' \Omega_{\Lambda}}}
\right ) \eeq

\noi with

\beq \label{2.2e} \lambda = 1 - \Omega_M - \Omega_{\Lambda} \eeq

\noi and

\bea \label{2.3e} \widehat{\sin} (x) &=& \sin (x) \ \ \ {\rm for} \
\lambda < 0 \nn \\ &= & x \quad \qquad {\rm for} \ \lambda = 0 \nn \\
&=& \sinh (x) \ \ {\rm for} \ \lambda > 0 \ .\eea

\noi For small $z$ one obtains approximately

\beq \label{2.4e} D(z) \cong {cz \over H_0} \left ( 1 + {z \over 2}
\left ( 1 + \Omega_{\Lambda} - {1 \over 2} \Omega_M \right ) \right ) +
{\cal O} (z^3)\ . \eeq

\noi Hence, apart from $H_0$, $D(z)$ depends on the two parameters
$\Omega_M$ and $\Omega_{\Lambda}$. Their separate determination
requires, however, the measurement of the terms of ${\cal O} (z^3)$ in
(\ref{2.1e}) or (\ref{2.4e}).\par

In order to determine the distance $D(z)$ of distant objects we need
objects with very well known absolute magnitude $M$. These are
supernovae of type Ia, which light up within a few weeks
and fade away within a few months. From the light curve the absolute
magnitude can be determined to high precision. \par

The luminosity distance $D$ then follows from a measurement of the
apparent magnitude $m$, which is related to $M$ and $D$ by

\beq \label{2.5e} m = M + 5 \log (D) + K \ , \eeq

\noi where $K$ includes a constant and a correction for the variation
of the apparent magnitude with the redshift. (The factor 5 and
the logarithm of $D$ in (\ref{2.5e}) have its origin in the logarithmic
scale for apparent and absolute magnitudes.) \par

The search for high redshift ($z \sim .4-.9$) supernovae starts with
the observation of patches of sky with tens of thousands of galaxies.
Some weeks later the same patches are observed again, and by comparison
one finds a few dozens of supernovae, which have not been there before.
Generally these have not yet reached peak brightness, and subsequently
their light curves are tracked more or less continuously, partly with
the help of the Hubble Space Telescope. \par

Two experimental groups lead by S. Perlmutter \cite{5r} and
B. Schmidt \cite{6r} have pursued this program, and results from best
fits to $D(z)$ up to $z\sim 1$ (one with $z\sim 1.7$ in \cite{6r}) can
directly be plotted in the plane $\Omega_{\Lambda}$ versus $\Omega_M$,
see fig. 1. The results of both groups agree well, and systematic
errors as absorption of light by dust, supernovae evolution and
selection bias are believed to be under control. \par

The figure shows that \par

-- $\Omega_{\Lambda} = 0$ is ruled out at the 99 percent level, assuming
$\Omega_M > 0$ (but no other assumptions on $\Omega_{\Lambda} +
\Omega_M$) \par

-- a flat universe ($k = 0)$, i.e. $\Omega_M +
\Omega_{\Lambda} = 1$, is consistent with the $1\sigma$ contour, and
leads to $\Omega_{\Lambda} \sim 0.7$, $\Omega_M \sim 0.3$. \par

The essential and amazing features of this result are \par

a) $\Omega_{\Lambda}$ is tiny, but non-vanishing: For the density
$\Lambda$ one obtains (with $c = 1$)

\beq \label{2.6e} \Lambda \cong 6 \cdot 10^{-24} \ {g \over m^3} \ .
\eeq

b) $\Omega_{\Lambda}$ is of the same order as $\Omega_M$, which is a
priori difficult to understand, since both quantities depend very
differently on the time: Whereas $\rho (t)$ decays like $t^{-2}$ (cf.
(\ref{1.35be})), $\Lambda$ remains constant, hence the ratio
$\Omega_M/\Omega_{\Lambda}$ also decays like  $t^{-2}$.\par

Hence we happen to live in an epoch where $\Omega_M$ and
$\Omega_{\Lambda}$ are of comparable order of magnitude; this is the
so-called ``coincidence problem''.

\mysection{The Cosmological Constant in Classical and Quantum Field
Theory} \hspace*{\parindent} As long as we consider classical general
relativity coupled to matter in the form of point-like sources (or a
fluid made out of point particles) the cosmological constant appears as
an arbitrary free parameter. Note that this framework is consistent
with all present tests of general relativity. \par

Problems appear once we extrapolate from length scales of ${\cal O}$(1
mm) (where gravity is tested) down to length scales of 1 Fermi, 1
TeV$^{-1}$ or even smaller, where we describe matter by classical or
quantum field theory. Usually it is assumed that the consistent
coupling of macroscopic matter to gravity has its origin in microscopic
Lagrangians (for fundamental fields describing elementary particles)
which are invariant under general coordinate transformations, cf. the
Lagrangian for a scalar field in chapter 1.\par

The Higgs sector of the standard model is traditionally described by
such a Lagrangian for the Higgs field $H$, including a potential

\beq \label{3.1e} V^H(H^2) \cong V_0^H + {\cal O}(H^2) \eeq

\noi where we have developed $V^H$ around its non-trivial minimum. The
correct dimension of the constant $V_0^H$ is $g/$(m sec$^2$), which is
obtained from the condition that the action

\beq \label{3.2e} S = {1 \over \hbar} \int dt \ d^3x \ V_0^H \eeq

\noi is dimensionless. (Recall that $\hbar \cong 10^{-31} gm^2$/sec.)
This dimension coincides with the dimension of $\Lambda$ as obtained
from eq. (\ref{1.44e}) with $\Omega_{\Lambda}$ dimensionless, and the
dimensions of $H_0$ and $\kappa$ given in eqs. (\ref{1.37e}) and
(\ref{1.4e}). Traditionally, however, particle physicists put $c =
\hbar = 1$. In these units the natural scale of $V_0^H$ is simply given
by the 4th power of the Higgs vev, i.e.

\beq \label{3.3e} V_0^H \cong (100 \ {\rm GeV})^4 = 10^8 \ {\rm GeV}^4
\ . \eeq

\noi In order to bring eq. (\ref{3.3e}) into the form of eq.
(\ref{2.6e}) (which has the dimension of a matter density)
we have to use

\beq \label{3.4e} 1\ {\rm GeV}^4 \cong 2.3 \cdot 10^{23} \ {g \over
m^3} \ , \eeq

\noi hence (\ref{3.3e}) corresponds to a natural value of the
cosmological constant $\Lambda^H$ of the order

\beq \label{3.5e} \Lambda^H \sim 2 \cdot 10^{31} \ {g \over m^3} \ .
\eeq

\noi This is 55 orders of magnitude off the observed value
(\ref{2.6e}). \par

 From a Higgs sector of a grand unified theory, where

\beq \label{3.6e}
V_0^{GUT} \cong \left ( 10^{16} \ {\rm GeV}\right )^4 \ ,
\eeq

\noi we would obtain

\beq \label{3.7e} \Lambda^{GUT} \sim 2 \cdot 10^{87} \ {g \over m^3}
\ , \eeq

\noi which is 111 orders of magnitude away from (\ref{2.6e}). \par

What about quantum field theory? We cannot consistently quantize
gravity, but we can couple the metric $g_{\mu\nu}(x)$ as an external
field to the bare action of a quantum field theory such that it becomes
invariant under general coordinate transformations. For its partition
function we write schematically

\beq \label{3.8e} e^{i{G}({J},\ g_{\mu\nu})} = \int {\cal D}
\varphi e^{iS_{bare} (\varphi,\ g_{\mu\nu}) + i {J} \cdot \varphi
} \eeq

\noi where $\varphi$ are quantum fields and ${J}$ the
corresponding sources. Then $g_{\mu\nu}$ plays actually the role of a
source for the (composite) energy-momentum operator, i.e. functional
derivatives of (\ref{3.8e}) with respect to $g_{\mu\nu}$ generate the
corresponding matrix elements. Instead of the functional ${G}
({J}, g_{\mu\nu})$ in (\ref{3.8e}) it is more convenient to work
with the effective action $\Gamma_{eff}(\varphi_{c\ell}, g_{\mu\nu})$,
which is obtained by a Legendre transform from ${G}({J},
g_{\mu\nu})$, keeping $g_{\mu\nu}$ fixed. Now the components of the
energy momentum tensor, which have to be inserted into the Einstein
equations (\ref{1.3e}), are obtained in analogy to eq. (\ref{1.7e}):

\beq \label{3.9e} T_{\mu\nu}(x) = {2 \over \sqrt{-g(x)}} \ {\delta
\over \delta g^{\mu\nu}(x)} \ \Gamma_{eff} (\varphi_{c\ell},
g_{\mu\nu})  \eeq

\noi where, in the vacuum, the fields $\varphi_{c\ell}$ are extrema of 
$\Gamma_{eff} (\varphi_{c\ell}, g_{\mu\nu})$. The leading diagrams,
which contribute to the diagonal components of $T_{\mu\nu}$ and hence
to the cosmological constant, are quartically (!) divergent tadpole
diagrams. Hence an UV cutoff $\Lambda^{UV}$ is required, and the
quantum contribution $\Lambda^{Quant}$ to the cosmological constant is
of the order

\beq \label{3.10e} \Lambda^{Quant} \sim {1 \over 16 \pi^2} \left (
\Lambda^{UV}\right )^4 \ . \eeq

\noi If one uses the Planck scale $\sim 10^{19}$~GeV as an UV cutoff
(under the assumption that local quantum field theory is valid up to
this scale) eq. (\ref{3.10e}) gives

\beq \label{3.11e} \Lambda^{Quant} \sim 10^{97} {g \over m^3}
\eeq

\noi which is 120 orders of magnitude off the experimental result
(\ref{2.6e}). This is the world record on disagreement between theory
and experiment. \par

What size of the UV cutoff $\Lambda^{UV}$ could we tolerate in order
not to be in conflict with the experimental result (\ref{2.6e})? A
rapid calculation show that we would need

\beq \label{3.12e} \Lambda^{UV} \ \lsim \ 8 \cdot 10^{-3}\ {\rm eV}
\eeq

\noi or

\beq \label{3.13e} \left ( \Lambda^{UV}\right )^{-1} \ \gsim \ 0.025 \
{\rm mm} \ . \eeq

Hence even very low energy quantum effects, at sizes much larger than
the size of atoms (where quantum mechanics is well tested), lead to
untolerable contributions to the cosmological constant. \par

The contradiction between the field theoretical results for
$\Lambda^H$, $\Lambda^{GUT}$ and $\Lambda^{Quant}$ and the
experimental result (2.6) is the famous "cosmological constant problem"
\cite{ccr}.

\newpage

\mysection{Towards Solutions of the Problem}

\noi {\bf a) \underbar{Supersymmetry}} \\

Unbroken global supersymmetry \citt{7r}{8r}{9r} would lead to a nice
solution of the cosmological constant problem. This is because the
energy operator $P^0$ can be written as a sum over squares of
supercharges $Q_{\alpha}$:

\beq \label{4.1e} P^0 = {1 \over 4} \sum_{\alpha} \left ( Q_{\alpha} +
\overline{Q}_{\alpha} \right ) \left ( Q_{\alpha} +
\overline{Q}_{\alpha} \right ) \ , \eeq

\noi hence matrix elements of $P^0$ can be written as

\beq \label{4.2e} <\psi |P^0|\psi > = {1 \over 4} \sum_{\alpha}
<\psi_{\alpha} | \psi_{\alpha}> \eeq

\noi with

\beq \label{4.3e} \psi_{\alpha} = \left ( Q_{\alpha} +
\overline{Q}_{\alpha} \right ) | \psi > \ . \eeq

\noi The right-hand side of eq. (\ref{4.2e}), and hence all expectation
values of $P^0$, is semi-positive in a semi-positive Hilbert space.
Thus, if a single state $\psi_0$ with $<\psi_0 |P^0|\psi_0> = 0$
exists, it is the state of lowest (vanishing) energy and hence the
vacuum state. This argument includes all quantum effects, and would
nicely explain a vanishing energy in the vacuum. However, eqs.
(\ref{4.2e}) and (\ref{4.3e}) show that in this vacuum state
supersymmetry would be unbroken,

\beq \label{4.4e} \left ( Q_{\alpha} + \overline{Q}_{\alpha} \right ) |
\psi_0 > = 0 \eeq

\noi which would imply untolerable Fermion/Boson degeneracies in the
physical spectrum. \par

In classical $N = 1$ supergravity it is possible to break spontaneously
global supersymmetry, with vanishing vacuum energy and without fine
tuning, provided the kinetic Lagrangian of the $n$ matter fields is
invariant under a non-linearly realized global $SU(n,1)$ symmetry, i.e.
if it parametrizes the coset space of a $SU(n,1)/SU(n) \times U(1)$
non-linear sigma model \citd{9r}{10r}.
Such $N=1$ supergravity theories in $d= 4$ appear naturally by
compactifying $N=1$ supergravity in $d = 10$ down to $d = 4$, hence in
most realistic superstring theories to lowest order in $\alpha'$ 
\cite{11r}.\par

However, here the vanishing of the vacuum energy is a purely classical
phenomenon and spoiled by quantum corrections. If the spontaneous
breakdown of supersymmetry manifests itself in the form of gaugino
masses, positive definite scalar masses or trilinear scalar couplings
of order $M_{susy}$, the quantum corrections to the cosmological
constant are still of the order

\beq \label{4.5e} {1 \over 16 \pi^2} \ M_{susy}^2 \left ( \Lambda^{UV}
\right )^2 \ ,  \eeq

\noi which is much too large for $M_{susy} \sim 100$~GeV and
$\Lambda^{UV} \sim M_{Planck}\sim 10^{19}$~GeV. \par

If the spontaneous breakdown of supersymmetry manifests itself
{\it only} in the form of opposite mass shifts among scalars and
pseudo-scalars of a massive chiral multiplet (so-called $F$-type
splitting) \citt{9r}{10r}{12r} the quantum corrections to the 
cosmological constant are reduced to the order

\beq \label{4.6e} {1 \over 16 \pi^2} \ M_{susy}^4 \eeq

\noi to all orders in perturbation theory. However, this value is still
too large, and comparable to the contribution of the classical Higgs
potential in the classical standard model. Hence, since a supersymmetry
breaking scale $M_{susy} \ \lsim \ 10^{-2}$~eV (cf. eq. (\ref{3.12e}))
is out of question, supersymmetry alone is finally not able to explain
the measured value of the cosmological constant. \\

\noi {\bf b) \underbar{Dilaton}} \\

It is tempting to obtain a vanishing vacuum energy as a consequence of
the equation of motion of a scalar field. A natural candidate for such
a field is the dilaton $\phi$, which helps to realize scale invariance
non-linearly: In the presence of a dilaton all mass scales $M$ appear
multiplied with an exponential of $\lambda\phi$. (The smallest possible
value for the coupling $\lambda$ is $\lambda\sim \sqrt{\cal{G}} =
{M_{Planck}}^{-1}$.) Hence the ``vacuum energy'' $V_0(\phi )$, which is
of the order $M^4$, is roughly of the form

\beq \label{4.7e} V_0(\phi ) \sim M^4 e^{4\lambda\phi} \ . \eeq

\noi Scalars with this property appear in string theory and
compactified supergravity theories; often the dilaton $\phi$ with the
property (\ref{4.7e}) in $d = 4$ is a linear combination of several
such scalars. \par

At first sight eq. (\ref{4.7e}) implies indeed

\beq \label{4.8e} {dV_0(\phi ) \over d \phi} = 0 \leftrightarrow
V_0(\phi ) = 0 \ .\eeq

\noi The value $\phi_0$ which solves (\ref{4.8e}) is obviously
$\phi_0 = - \infty$. This is the ``dilaton run away problem'': In
string theory and compactified supergravity theories couplings depend
typically on $\phi$, and these couplings vanish (or even tend to
infinity in some cases) in this limit. \par

Moreover particle masses $m$ depend necessarily on the dilaton in the
form $m(\phi ) \sim M e^{\lambda\phi}$. Hence, whenever $V_0(\phi )$
vanishes, all masses vanish as well (and scale invariance is restored).
In addition, near the minimum of its potential the dilaton itself is
nearly massless, and the coupling of the dilaton to particles of mass
$m$ is of the order of $\sqrt{\cal{G}} \cdot m = m/M_{Planck}$. Light
scalars induce long-range interactions which are strongly constraint by
limits on "fifth forces" and/or violations of the equivalence
principle. \par

Note that these arguments apply also if mass scales are generated by
dimensional transmutation (as in QCD or technicolour) once the $\phi$
dependence of the UV cutoff is correctly taken into account. Hence,
despite many efforts in this direction \cite{ccr}, no working model
could be constructed up to date. \\

\noi  {\bf c) \underbar{3-Form-Field}} \\

A 3-form-field $A_{\mu\nu\rho} = A_{[\mu\nu\rho]}$ appears in $N = 8$
supergravity \cite{14r}. It has a field strength $F_{\mu\nu\rho\sigma}
= \partial_{[\mu} A_{\nu\rho\sigma]}$ and satisfies the equations of
motion

\beq \label{4.9e} \partial^{\mu} F_{\mu\nu\rho\sigma} = 0 \eeq

\noi (suitably covariantized). The equation of motion (\ref{4.9e})
follows from a Lagrangian

\beq \label{4.10e} {\cal L}_A = \lambda \ F_{\mu\nu\rho\sigma} \
F^{\mu\nu\rho\sigma} \ . \eeq

\noi The only non-trivial solutions of (\ref{4.9e}) which respect
Lorentz covariance are of the form

\beq \label{4.11e} F_{\mu\nu\rho\sigma} = \Sigma \
\varepsilon_{\mu\nu\rho\sigma} \eeq

\noi where $\Sigma$ is an arbitrary constant. If one naively plugs
(\ref{4.11e}) back into ${\cal L}_A$, one obtains an effective
contribution to the cosmological constant

\beq \label{4.12e} V_A = - 24 \ \lambda \ \Sigma^2 \ .\eeq

\noi This does not solve the cosmological constant problem, but the
arbitrary value of $\Sigma$ could be used to cancel other contributions
to it. Based on a path integral approach to Euclidean quantum gravity
Hawking has argued \cite{15r} that, if $S_{Eucl}$ contains an arbitrary
parameter as in (\ref{4.12e}), the most probable value of $\exp
(-S_{Eucl})$ is where $S_{Eucl}$ is minimal. Chosing the four sphere
$S^4$ for Euclidean space time, the minimum of $S_{Eucl}$ (actually
$-\infty$) is achieved for a vanishing cosmological constant, i.e.
infinite volume. Subsequently Duff has pointed out \cite{16r}, however,
that one should not plug ans\"atze for solutions back into the action,
but rather vary the unconstrained action, and that consequently the
$A$-contribution to the Euclidean action is positive, invalidating
Hawking's argument.\par

Brown and Teitelboim \cite{17r} have coupled a $d-1$-form field $A$
(in $d$ space-time dimensions) to a $d-2$-brane. Based on a
2-dimensional toy model they proposed that then the value of $\Sigma$
could relax dynamically to the one corresponding to a vanishing
cosmological constant. \par

But, all in all no convincing mechanism is known which naturally
generates such a value of $\Sigma$ in $d = 4$.\\

\noi  {\bf d) \underbar{Quintessence}} \\

Quintessence \cite{18r} does not aim at a solution of the problem of
the smallness of the cosmological constant with respect to particle
physics scales, but at a solution of the coincidence problem: Why is
$\Omega_{\Lambda}$ of the order of $\Omega_M$?\par

In this approach the cosmological constant $\Lambda$ is replaced by a
potential $V(\phi )$ of a scalar field $\phi$, which is {\it
not} assumed to have reached its minimum value by now. One uses that
for potentials with

\beq \label{4.13e} {V''V \over (V')^2} \geq 1 \eeq

\noi the values of the scalar fields $\phi$ evolve with time such that,
for a wide range of initial conditions, $V(\phi )$ is of the order of
the density of background (standard) matter today. The solution of the
coupled Einstein equations and equations of motion for $\phi$ is
required for this result. As a by-product one finds that the effective
parameter $w_{\phi}$ (cf. eq. (\ref{1.16e}) ff), associated to the
equation of state for $\phi$, is often time-dependent. \par

A simple potential satisfying (\ref{4.13e}) is given by \citd{19r}{20r}

\beq \label{4.14e} V(\phi ) \sim e^{-\lambda \phi} \ . \eeq

\noi However, now the full cosmological evolution has to be
reconsidered, and notably nucleosynthesis constraints (which require
the quintessence energy density not to be too large at that epoch) rule
out the simple model (\ref{4.14e}). Among the many working models is
\cite{19r}

\beq \label{4.15} V(\phi ) \sim {\lambda \over \phi^{\alpha}} \quad ,
\quad 0 < \alpha \ \lsim \ 2 \ . \eeq

Motivations for quintessence fields and potentials can again be found
in dilaton (or moduli) sectors of superstring or supergravity models,
where the ``dilaton run away problem'' is now turned into a goody.
However, the coupling of $\phi$ to  matter has to be reduced ad hoc in
order not to generate additional dangerous long-range forces.
Moreover, of course, the absolute value of the cosmological constant
(the minimum of all scalar potentials) is still fine-tuned to zero
here; the finite observed value of $\Omega_{\Lambda}$ is just explained
by the fact that the quintessence field $\phi$ has not yet reached the
minimum of its potential.

It should be noted that once $D(z)$ (cf. (\ref{2.1e})) can be determined
to even better accuracy by measuring even more supernovae, various
quintessence scenarios, i.e. various dependences of the cosmological
"constant" on time, can be distinguished experimentally.\\

\noi  {\bf e) \underbar{Brane Universes}} \\

In brane worlds the dimension of space-time is extended beyond $d = 4$.
In contrast to the standard Kaluza-Klein approach, however, matter
fields are confined to live on $3 + 1$ dimensional manifolds (3-branes)
which are embedded into the higher dimensional space-time manifold.
Only gravity (and so-called bulk fields) lives in this higher
dimensional manifold. \par

The most studied examples consist in one extra fifth dimension (denoted
by $y$) and one or two embedded 3-branes. The fifth dimension is
compact, i.e. $y$ varies between $-\pi R_5 \leq y \leq + \pi R_5$ where
the points $y = \pm \pi R_5$ are identified. The branes are located at
$y = 0$ (and $y = \pm \pi R_5$), and only symmetric modes of the
$y$-dependent metric $g_{\mu\nu}(x, y)$ under $y \leftrightarrow - y$
are allowed. \par

This example is motivated by the Ho\v rava-Witten construction of the
strong coupling limit of the heterotic superstring \cite{21r}, which is
formulated in 11 dimensions with two 9-branes. After compactification
of 6 space dimensions the above picture in 5 dimension emerges. Many
more general brane worlds have been constructed since then, motivated
by the presence of Dirichlet branes in string theories \cite{22r}. \par

In this approach one can put constant energy densities $\Lambda_1$,
$\Lambda_2$ on either of the two branes, and another constant energy
density $\Lambda_b$ in the bulk outside the branes. Which of these
plays the role of the cosmological constant in the
Friedmann-Robertson-Walker cosmology (\ref{1.33e})? \par

In order to answer this question one has to start with the five
dimensional Einstein equations for $g_{\mu\nu}(x, y)$.  Assuming
translational invariance in the three spacial $x^i$ dimensions and a
constant diagonal energy momentum tensor $T_{\mu}^{\ \nu}$ in the bulk
the $y$ dependence of $g_{\mu\nu}$ can be fixed completely. Actually
the first derivatives of $g_{\mu\nu}$ with respect to $y$ are
discontinuous across the branes. This solution for the $y$ dependence
of $g_{\mu\nu}$ can be plugged back into the Einstein equations, and
the $t$ dependence of $g_{\mu\nu}$ on the brane(s) can be parametrized
by a Friedmann-Robertson-Walker ansatz (\ref{1.12e}). \par

Although the five-dimensional Einstein equations contain additional
terms compared to the four dimensional Einstein equations (involving
non-vanishing terms $\sim \partial_y g_{\mu\nu}$) the resulting
equations for $a(t)$ can be written in the form of eq. (\ref{1.33e})
plus corrections on its right-hand side \cite{23r}, provided $\Lambda_2
= - \Lambda_1$ (as in the Ho\v rava-Witten construction). The term
$\kappa \Lambda$ in eq. (\ref{1.33e}) is now replaced by

\beq \label{4.16e} \kappa \ \Lambda_{eff} = {\kappa_5 \over 2} \
\Lambda_b + {\kappa_5^2 \over 12} \ \Lambda_1^2\eeq

\noi where $\kappa_5$ is the five dimensional gravitational coupling.

More generally, in the presence of bulk fields $\varphi_i$ with
potentials $V^1(\varphi ) = - V^2(\varphi )$ on the branes, a potential
$V_b(\varphi )$ in the bulk and a general sigma model metric ${\cal
G}_{ij}(\varphi )$ for the kinetic terms in the bulk, $\Lambda_{eff}$
corresponds to the minimum of the effective potential \cite{24r}

\beq \label{4.17e} V_{eff}(\varphi ) = {1 \over 2}V_b - {1 \over 32} \
V^1_{,i} \ {\cal G}^{ij} \ V_{,j}^1 + {\kappa_5 \over 12} (V^1)^2\
.\eeq

\noi (Curiously enough $V_{eff}$ ressembles the scalar potential in $N
= 1$ supergravity, with $V^1$ playing the role of the superpotential).
\par

Generically even two five tunings ($\Lambda_2 = - \Lambda_1$, and the
vanishing of the right-hand side of eq. (\ref{4.16e}) through an
appropriate choice of $\Lambda_b$) are required in order to reproduce
both a small effective cosmological constant and a time independent
effective four dimensional gravitational constant in brane worlds with
two 3-branes. Scenarios with just one 3-brane and a non-compact fifth
dimension also exist \cite{25r} where just one fine-tuning is required.
\par

It has been claimed that even this fine-tuning can be avoided if one
puts a scalar field in the bulk, with either no potential at all in the
bulk and an exponential potential on the brane \cite{26r} or an
exponential potential in the bulk \cite{27r}: Then the combined
equations of motion for the metric and the scalar field generically
possess a (static) solution which corresponds to a vanishing effective
cosmological constant. This scenario seems to violate the above theorem
(4.17) on the effective cosmological constant. Indeed this
``self-tuning''-scenario always involves a $y$-dependent metric with
naked singularities in $y$. As shown in \cite{28r}, any attempt to
regularize these singularities requires a new fine-tuning. Moreover
this scenario would require fine-tuned initial conditions. \par

The previous scenario involving a 3-form-field has also been applied to
brane universes \cite{29r}. However, again fine-tuned initial
conditions are required if the static solution corresponding to a
vanishing effective cosmological constant is to be realized \cite{30r}.

\mysection{Conclusions} \hspace*{\parindent}

None of the attempts listed above has led to a successful explanation
for the observed smallness of the cosmological constant. We did not
have time to discuss the approach of Verlinde et al. \cite{31r}, which
is based on an interpetation of the AdS/CFT correspondence as an
AdS/Renormalization Group correspondence. This concept is reviewed in
\cite{32r}. \par

Let us recall that the origin of the problem is the coupling of local
four dimensional quantum field theory (supposedly valid down to length
scales of ${\cal O}$((100 GeV)$^{-1}$)) to gravity, whose Einstein action
is tested down to length scales of ${\cal O}$(1 mm). Any attempt to
solve the cosmological constant problem -- involving extra dimensions,
branes, AdS/CFT or whatsoever -- must either fit into these frameworks
(possibly with new extra fields) or be precise on its modifications.
\par

Modifications of the Einstein action -- either through effects due to
quantum gravity, ``large'' extra dimensions, curvature squared terms
from string theory etc. -- face the following obstacle: Gravitational
interactions from millimeters to astronomical distances should not be
(drastically) modified, but there should be an effect on the equation
of motion of the ``global'' Robertson-Walker mode $a(t)$ of the
gravitational field -- without spoiling the very successful part of the
cosmological standard model. Notably no ``short-distance''
modifications (affecting only small wave length modes of the
gravitational field) can do this job. \par

Recall also that extra dimensions -- with or without branes -- can
{\it always} be represented in terms of effective four dimensional
fields (including possibly infinite towers of massive states) with
local interactions. Hence any cancellation of the vacuum energy within
an approach based on extra dimansions must be
representable by an effective four-dimensional Lagrangian with peculiar
properties. \par

Then one finds oneself automatically in one of the frameworks (dilaton
etc.) considered and discarded before. It seems that any dynamics
sensitive to a vacuum energy density of ${\cal
O}$(($10^{-3}$~eV)$^{4}$) must involve fields with masses of this order
or lighter. On the one hand these fields have to couple to matter
and/or gravity (in order to detect the vacuum energy), but they should
neither imply new long-range interactions nor have disastrous
cosmological effects like important relic densities. Altogether these
constraints seem to be self-contradictory. \par

In this situation some scientists appeal to anthropic principles as,
e.g., ``we happen to live in one long-living among $\sim 10^{100}$
possible universes''. Otherwise we possibly have to touch at one of the
``hidden assumptions'' -- the local coupling of fundamental fields to
gravity. Work in this direction is in progress \cite{33r}.

\vskip 1cm

{\Large \bf Acknowledgement}
\vskip 1cm

I would like to  thank the participants of the Workshop for various
critical and helpful comments on the subject.

\newpage 

\def\labelenumi{[\arabic{enumi}]} \noindent {\large\bf
References} \ben

\item\label{1r} S. Weinberg, {\it Gravitation and Cosmology}, John 
Wiley \& Sons, New York, 1972.

\item\label{2r} R. Sexl, H. Urbantke, {\it Gravitation and 
Kosmologie}, B I Mannheim, 1983.

\item\label{3r} R. Wald, {\it General Relativity}, The University of 
Chicago Press, Chicago, 1984.

\item\label{4r} A. Krasi\'nski, {\it Inhomogeneous Cosmological 
Models}, Cambridge University Press, Cambridge (UK), 1997.

\item\label{5r} S. Perlmutter et al., Int. J. Mod. Phys. {\bf A15 S1} 
(2000) 715, eConf/990809, Astrophys. J. {\bf 517} (1999) 565.

\item\label{6r} A. Riess et al., Astron. J. {\bf 516} (1998) 1009, 
Astrophys. J. {\bf 560} (2001) 49 (astro-ph/0104455).

\item\label{cmbr} P. de Bernardis et al., Nature {\bf 404} (2000) 955.

\item\label{ccr} For reviews and references see S. Weinberg, Rev. 
Mod. Phys. {\bf 61} (1989) 1, and S. M. Carroll, astro-ph/0004075.

\item\label{7r}J. Wess, J. Bagger, {\it Supersymmetry and 
Supergravity}, Princeton University Press, Princeton, 1983.

\item\label{8r} H. P. Nilles, Phys. Rept. {\bf 110C} (1984) 1.

\item\label{9r} N. Dragon, U. Ellwanger, M. G. Schmidt, Progress in 
Part. and Nucl. Phys. {\bf 18} (1987) 1.

\item\label{10r} N. Dragon, U. Ellwanger, M. G. Schmidt, Phys. Lett. 
{\bf B145} (1984) 192; Nucl. Phys. {\bf B255} (1985) 549; Phys. 
Lett. {\bf B154} (1985) 373.

\item\label{11r} E. Witten, Phys. Lett. {\bf B155} (1985) 151;\\
U. Ellwanger, M. G. Schmidt, Nucl. Phys. {\bf B294} (1987) 445.

\item\label{12r} U. Ellwanger, Phys. Lett. {\bf B349} (1995) 57.

\item\label{14r} A. Aurilia, H. Nicolai, P. Townsend, Nucl. Phys. 
{\bf B176} (1980) 509.
 
\item\label{15r} S. Hawking, Phys. Lett. {\bf B134} (1984) 403.

\item\label{16r} M. Duff, Phys. Lett. {\bf B226} (1989) 36.

\item\label{17r} J. Brown, C. Teitelboim, Phys. Lett. {\bf B195} 
(1987) 177, Nucl. Phys. {\bf B297} (1988) 787.

\item\label{18r} For reviews see P. Bin\'etruy, Int. J. Theor. Phys. 
{\bf 39} (2000) 1859, hep-ph/0005037, and V. Sahni, astro-ph/0202076.

\item\label{19r} B. Ratra, P. Peebles, Phys. Rev. {\bf D37} (1988) 3406.

\item\label{20r} C. Wetterich, Nucl. Phys. {\bf B302} (1988) 668;\\
P. Ferreira, M. Joyce, Phys. Rev. Lett. {\bf 79} (1997) 4740  
Phys. Rev. {\bf D58} (1998) 023503.

\item\label{21r} P. Ho\v rava, E. Witten, Nucl. Phys. {\bf 460} 
(1996) 506,  Nucl. Phys. {\bf B475} (1996) 94.

\item\label{22r} J. Polchinski, Tasi Lectures on $D$-Branes,
hep-th/9611050.

\item\label{23r} P. Bin\'etruy, C. Deffayet, U. Ellwanger, D. 
Langlois, Phys. Lett. {\bf B477} (2000) 285.

\item\label{24r} O. De Wolfe, D. Freedmann, S. Gubser, A. Karch, 
Phys. Rev. {\bf D62} (2000) 046008; \\
U. Ellwanger, Phys. Lett. {\bf B473} (2000) 233.

\item\label{25r} M. Gogberashvili, Mod. Phys. Lett. {\bf A14} (1999)
2025; \\ L. Randall, R. Sundrum, Phys. Rev. Lett. {\bf 83} (1999) 4670.

\item\label{26r} N. Arkani-Hamed, S. Dimopoulos, N. Kaloper, R. 
Sundrum, Phys. Lett. {\bf B480} (2000) 193.

\item\label{27r} S. Kachra, M. Schulz, E. Silverstein, Phys. Rev. 
{\bf D62} (2000) 045021.

\item\label{28r} S. F\"orste, Z. Lalak, S. Lavignac, H. P. Nilles, 
Phys. Lett. {\bf B481} (2000) 360,  JHEP {\bf 0009} (2000) 34.

\item\label{29r}J. Kim, B. Kyae, H. Lee, Phys. Rev. Lett. {\bf 86} 
(2001) 4223,  Nucl. Phys. {\bf B613} (2001) 306.

\item\label{30r} A. Medved, hep-th/0109180.

\item\label{31r} E. Verlinde, Class. Quant. Grav. {\bf 17} (2000)
1277;\\ E. Verlinde, H. Verlinde, JHEP {\bf 0005} (2000) 34.

\item\label{32r} U. Ellwanger, Lectures at the LPT Orsay,
hep-th/0009006.

\item\label{33r} U. Ellwanger, hep-th/0201163.
\een

\begin{figure}[htb] 
\centerline{\epsfxsize=12cm\epsfbox{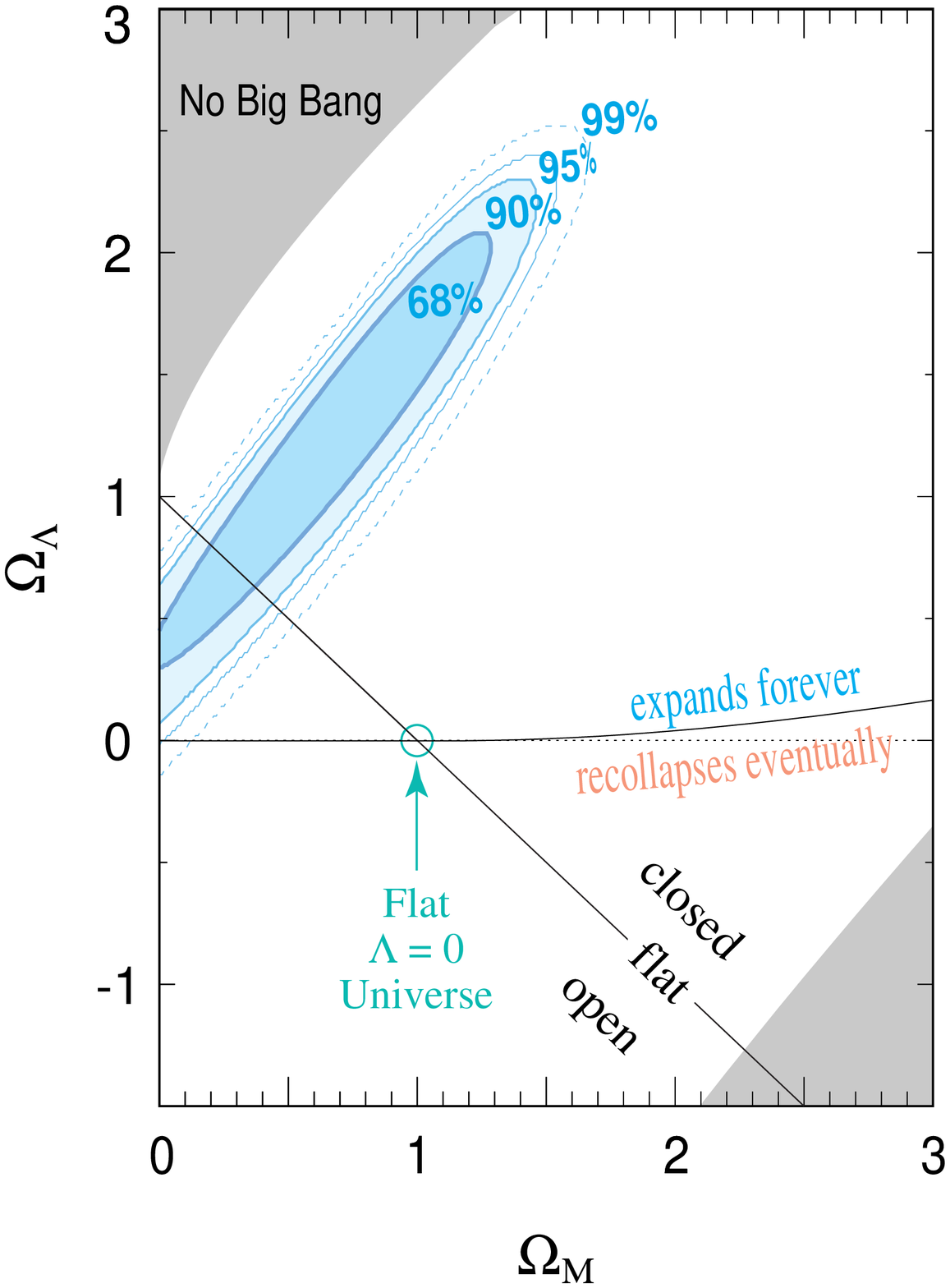}}
\caption{ Confidence regions in the $\Omega_M-\Omega_\Lambda$  plane
based on data from 42 type Ia supernovae at large redshift discovered
by the Supernova Cosmology Project \cite{5r}, and 18 supernovae at low
redshift.} 
\label{fig1} 
\end{figure}

\end{document}